\documentclass[12pt,a4paper]{article}
\usepackage{amssymb}
\usepackage{graphicx}
\usepackage{amsmath}
\usepackage{makeidx}
\usepackage{indentfirst}
\usepackage[T1]{fontenc}
\usepackage[utf8]{inputenc}
\usepackage{authblk}

\newcounter{resultnum}[section]
\newcounter{conclusionnum}[section]
\newcounter{conditionnum}[section]
\newcounter{conjecturenum}[section]
\newcounter{examplenum}[section]
\newcounter{exercisenum}[section]
\newcounter{lemmanum}[section]
\newcounter{notationnum}[section]
\newcounter{theoremnum}[section]
\newcounter{definitionnum}[section]
\newcounter{corollarynum}[section]
\newcounter{remarknum}[section]
\newcounter{propositionnum}[section]
\newcounter{acknowledgementnum}[section]
\newcounter{algorithmnum}[section]
\newcounter{axiomnum}[section]
\newcounter{casenum}[section]
\newcounter{claimnum}[section]
\newcounter{summarynum}[section]
\newcounter{problemnum}[section]
\begin{document}

\title{Equivalent Off--Diagonal Cosmological Models and Ekpyrotic Scenarios
in f(R)--Modified, Massive and Einstein Gravity}
\date{April 11, 2015}
\author{ \vspace{.1 in} {\large \textbf{Sergiu I. Vacaru}} \\
\vspace{.1 in} {\small \textit{University "Al. I. Cuza" Ia\c si, Rector's
Department, }}\\
{\small \textit{14 A. Lapu\c sneanu street, Corpus R, UAIC, office 323, Ia\c
si, Romania 700057 }}\\
{\small \textit{email: sergiu.vacaru@uaic.ro; sergiu.vacaru@gmail.com }} \\
{\small and}\\
{\small \textit{Theory Division, CERN, CH-1211, Geneva 23, Switzerland}%
\thanks{%
associated visiting research affiliation}} }
\maketitle

\begin{abstract}
We reinvestigate how generic off--diagonal cosmological solutions depending,
in general, on all spacetime coordinates can be constructed in massive and
f--modified gravity using the anholonomic frame deformation method. There
are constructed new classes of locally anisotropic and (in) homogeneous
cosmological metrics with open and closed spatial geometries. By resorting
such solutions, we show that they describe the late time acceleration due to
effective cosmological terms induced by nonlinear off--diagonal
interactions, possible modifications of the gravitational action and
graviton mass. The cosmological metrics and related St\" uckelberg fields
are constructed in explicit form up to nonholonomic frame transforms of the
Friedmann--Lama\^{\i}tre--Robertson--Walker (FLRW) coordinates. The
solutions include matter, graviton mass and other effective sources
modelling nonlinear gravitational and matter fields interactions with
polarization of physical constants and deformations of metrics, which may
explain dark energy and dark matter effects. However, we argue that it is
not obligatory always to modify gravity if we consider effective generalized
Einstein equations with nontrivial vacuum and/or non-minimal coupling with
matter. Indeed, we state certain conditions when such configurations mimic
interesting solutions in general relativity and modifications, for instance,
when we can extract the general Painlev\' e--Gullstrand and FLRW metrics. In
a more general context, we elaborate on a reconstruction procedure for
off--diagonal cosmological solutions which describe cyclic and ekpyrotic
universes. Finally, there are discussed open issues and further perspectives.

\vskip5pt

 \vskip5pt {\small %
}
\end{abstract}

\newpage

\tableofcontents


\section{Introduction}

The "ekpyrotic" and "new ekpyrotic" mechanisms and cyclic models have been
elaborated as alternatives to standard big bank inflationary cosmology \cite%
{kh}. Another alternative came from the idea that graviton may have a
nontrivial mass as it was proposed in the Fierz and Pauli work \cite{fierz}.
For recent reviews and related f(R) modifications and cosmological models,
in general, with non-minimal coupling and dilaton-brane cosmology, local
anisotropies and/or \ effective modelling of massive gravity, see
respectively \cite{odints}, \cite{kamensh} and \cite{kourets}. The modern
version of a ghost free (bimetric) massive gravity theory were made in a
series of papers (recent reviews can be found in \cite{hint}): \ Developing
generic nolinear versions of the Fierz-Pauli theory, the so--called vDVZ
discontinuity problem was solved via the Vainshtein mechanism \cite{dvz}
(avoiding discontinuity by going beyond the linear theory). There are also
more recent approaches based on DGP model \cite{dgp}. During a long time,
none solution was found for another problem with ghosts because at nonlinear
order in massive gravity appears a sixth scalar degree of freedom as a
ghost. This problem was cosndiered in a paper by Boulware and Deser \cite%
{boul} together with similar issues related to the effective field theory
approach.

A considerable amount of work has been made in order to undestand the
implications and find possible applications of of such ghost--free models.
The most substantial progress was made when de Rham and co--authors had
shown how to eliminate the scalar mode and Hassan and Rosen provided a
complete proof for a class of bigravity / bimetric gravity theories, see
\cite{drg}. In such approaches, the second metric describes an effective
exotic matter related to massive gravitons and does not suffer from ghost
instability to all orders in a perturbation theory and away from the
decoupling limit.

It is important to mention that the first black hole solutions in a
nonlinear massive theory were found in the context of high--energy physics
and for off--diagonal and/or higher dimensions generalizations \cite{bhnh}.
A general decoupling of generalized Einstein equations can be proven
following the anholonomic frame deformation methods, AFDM, \cite{vadm1}.
Here we also note that the possibility that the graviton has a nonzero mass $%
\mathring{\mu}$ results not only in fundamental theoretical implications but
give rise to straightforward phenomenological and cosmological consequences.
For instance, a gravitational potential of Yuka\-wa form $\sim e^{-\mathring{%
\mu}r}/r$ results in decay of gravitational interactions at scales $r\geq
\mathring{\mu}^{-1}$ and such an effect may result in the accelerated
expansion of the Universe. This way, a massive gravity theory, MGT, provides
alternatives to dark energy and, via effective polarizations of fundamental
physical constants may explain certain dark matter effects. We can treat
this as a result of generic off--diagonal nonlinear interactions. Recently,
various cosmological models derived for ghost free (modified) massive
gravity and bigravity theories have been elaborated and studied intensively
\cite{mgcosm,odints,kobayashi}.

The goal of this work is to construct generic off--diagonal cosmological
solutions in MGT and state the conditions when such configurations are
modelled equivalently in general relativity (GR). It extends the letter
variant of \cite{vlett} with new results and more details on proofs of
results and constructing exact solutions and further analysis of results and
speculation on reconstruction mechanism.

As a first step, we consider off--diagonal deformations of a "prime"
cosmological solution taken in general Painlev\'{e}--Gullstrand (PG) form,
when the Friedman--Lama\^{\i}tre--Robertson--Walker (FLRW) can be recast for
well--defined geometric conditions. Such constructions are performed in
section \ref{s2}.

At the second step, the "target" metrics will be generated to possess one
Killing symmetry (or other none Killing symmetries) and depend on timelike
and certain (all) spacelike coordinates. In general, such off--diagonal
solutions are with local anisotropy and inhomogeneities for effective
cosmological constants and polarizations of other physical constants and
coefficients of cosmological metrics which can be modelled both in MGT and
GR. We consider the method of constructing generic off--diagonal
cosmological solutions in section \ref{s3}.

Then (the third step), we shall emphasize and speculate on importance of
off--diagonal nonlinear gravitational interactions for elaborating
cosmological scenarios when dark matter and dark energy effects can be
explained by anisotropic polarizations of vacuum and/or de Sitter like
configurations. We provide examples of off--diagonal solutions with
solitonic configurations, see section \ref{s4}.

A reconstruction mechanism for off--diagonal cosmological solutions with
modified gravity and/or massive graviton effects is elaborated in section %
\ref{s5}. Finally, we conclude the paper in section \ref{sconcl}.

\section{Equivalent modelling of $f$--modified and massive gravity theories}

\label{s2}

We shall work with MGTs modelled on a pseudo--Riemannian spacetime $\mathbf{V%
}$ with physical metric $\mathbf{g=\{g}_{\mu \nu }\}$ and fiducial metrics.
On massive gravity, see reviews \cite{hint} and on geometric methods in
gravity and constructing exact solutions, see Refs. \cite{vadm1} . In
addition to well--known approaches with diadic and tetradic (vierbein)
variables, we shall work with nonholonomic manifolds when certain classes of
frame transforms can be adapted to preserve a chosen splitting of the
nonlinear and linear connection structures into some standard components
(for instance, defining the Levi Civita, LC, connection) and distortion
tensors which can be fixed to be zero if additional constraints are imposed.

Using different nonholonomic frame variables, the action for our model can
be written in two forms,
\begin{eqnarray}
S &=&\frac{1}{16\pi }\int \delta u^{4}\sqrt{|\mathbf{g}_{\alpha \beta }|}[%
\widehat{f}(\widehat{R})-\frac{\mathring{\mu}^{2}}{4}\mathcal{U}(\mathbf{g}%
_{\mu \nu },\mathbf{K}_{\alpha \beta })+\ ^{m}L]  \label{act} \\
&=&\frac{1}{16\pi }\int \delta u^{4}\sqrt{|\mathbf{g}_{\alpha \beta }|}%
[f(R)+\ ^{m}L].  \label{act1}
\end{eqnarray}%
In above formulas, the physical/geometrical objects with "hat" and/or
written in "boldface" form are considered for a conventional $2+2$ splitting
when the two dimensional horizontal, h, and the two dimensional vertical, v,
coordinates are labelled (respectively) in the form $u^{\alpha
}=(x^{i},y^{a}),$ or $u=(x,y),$ with indices $i,j,k...=1,2$ and $a,b,...=3,4.
$ The scalar curvature $R$ is for the LC--connection $\nabla$.

We write by $\widehat{R}$ the scalar curvature for an auxiliary (canonical)
connection $\widehat{\mathbf{D}}$ uniquely determined by two conditions:

\begin{enumerate}
\item it is metric compatible, $\widehat{\mathbf{D}}\mathbf{g}=0, $ and

\item the $h$- and $v$-torsions are zero (but there are nonzero $h-v$
components of torsion $\widehat{\mathcal{T}}$ completely determined by $%
\mathbf{g)}$ for a conventional splitting
\begin{equation*}
\mathbf{N}:\ T\mathbf{V=}h\mathbf{V\oplus }v\mathbf{V,}
\end{equation*}
with local coefficients of the so--called nonlinear connection,
N--connection structure, labeled in the form $\mathbf{N}= \{N_{i}^{b}\}.$
\end{enumerate}

General frame transforms can be parameterized in the form $e_{\alpha }=A_{\
\alpha }^{\alpha ^{\prime }}(u)\partial _{\alpha ^{\prime }},$ where the
matix $A_{\ \alpha }^{\alpha ^{\prime }}$ is non--degenerate in a finite, or
infinite region of $\mathbf{V}$ and $\partial _{\alpha ^{\prime }}=\partial
/\partial u^{\alpha ^{\prime }}.$ Using such $A_{\ \alpha }^{\alpha ^{\prime
}},$ we can always re--define the geometric and physical object with respect
to a class of N--adapted (dual) bases
\begin{eqnarray}
\mathbf{e}_{\alpha } &=&(\mathbf{e}_{i}=\partial _{i}-N_{i}^{b}\partial
_{b},e_{a}=\partial _{a}=\partial /\partial y^{a})\mbox{ and }  \notag \\
\mathbf{e}^{\beta } &=&(e^{j}=dx^{i},\mathbf{e}^{b}=dy^{b}+N_{c}^{b}dy^{c}),
\label{nframes}
\end{eqnarray}%
which are nonholonomic (equivalently, anholonomic) because, in general,
there are satisfied relations of type
\begin{equation*}
\mathbf{e}_{\alpha }\mathbf{e}_{\beta }-\mathbf{e}_{\beta }\mathbf{e}%
_{\alpha }=W_{\alpha \beta }^{\gamma }\mathbf{e}_{\gamma },
\end{equation*}%
for certain nontrivial anholonomy coefficients $W_{\alpha \beta }^{\gamma
}(u).$ The Einstein summation rule on repeating indices will be applied if
the contrary is not stated.

The connection $\widehat{\mathbf{D}}$ allows us to decouple the field
equations in various gravity theories and construct exact solutions in very
general forms. Here we note that the distortion relation from the LC
connection
\begin{equation*}
\widehat{\mathbf{D}}=\nabla +\widehat{\mathbf{Z}}[\widehat{\mathcal{T}}]
\end{equation*}%
is uniquely determined by a distorting tensor $\widehat{\mathbf{Z}}$
completely defined by $\widehat{\mathcal{T}}$ and (as a consequence for such
models) by $(\mathbf{g},\mathbf{N}).$ The main idea of the AFDM \cite{vadm1}
is to use $\widehat{\mathbf{D}}$ as an auxiliary [for the (pseudo)
Riemennian spacetimes] and/or [with nonholonomically induced torsion]
connection which positively allows us to decouple gravitational field
equations for very general conditions with respect to N--adapted frames (\ref%
{nframes}). We can not perform such a decoupling in general form if we work
from the very beginning with the data $(\nabla ,\partial _{\alpha ^{\prime
}})$ but there are proofs that this is possible for $(\widehat{\mathbf{D}},%
\mathbf{e}_{\alpha };\mathbf{g},\mathbf{N}).$ Having constructed integral
varieties (for instance, locally anisotropic and/or inhomogeneous
cosmological ones), we can impose additional nonholonomic (non--integrable
constraints) when $\widehat{\mathbf{D}}_{\mid \widehat{\mathcal{T}}%
=0}\rightarrow \nabla $ and $\widehat{R}\rightarrow R,$ where $R$ is the
scalar curvature of $\nabla ,$ and it is possible to extract exact solutions
in GR.\footnote{%
There will be considered also left and right up/low indices as labels for
some geometric/physical objects.}

The MGTs with actions of type (\ref{act}) generalize the so--called modified
$f(R)$ gravity and the ghost--free massive gravity \cite{drg}. We shall
follow some conventions from \cite{kobayashi}. There will be used the units
when $\hbar =c=1$ and the Planck mass $M_{Pl}$ is defined via $%
M_{Pl}^{2}=1/8\pi G,$ with 4--d Newton constant $G.$ We write $\delta u^{4}$
instead of $d^{4}u$ because there are used N--elongated differentials as in (%
\ref{nframes}). A model can be specified by corresponding constants,
dynamical physical equations and their solutions in corresponding variables.
It will be considered that $\mathring{\mu}=const$ is the mass of graviton.
For LC--configurations, we can fix conditions of type
\begin{equation}
\widehat{f}(\widehat{R})-\frac{\mathring{\mu}^{2}}{4}\mathcal{U}(\mathbf{g}%
_{\mu \nu },\mathbf{K}_{\alpha \beta })=f(\widehat{R}),\mbox{ \ or \ }%
\widehat{f}(\widehat{R})=f(R),\mbox{ \ or \ }\widehat{f}(\widehat{R})=R,
\label{mgrfunct}
\end{equation}%
which depend on the type of models we elaborate and what classes of
solutions we wont to construct. Such conditions can be very general ones for
arbitrary frame transforms and N--connection deformations. We emphasize that
it is possible to find solutions in explicit form if we chose the
coefficients $\{N_{i}^{a}\}$ and the local frames for $\widehat{\mathbf{D}}$
when $\widehat{R}=const$ in such forms that $\partial _{\alpha }\widehat{f}(%
\widehat{R})=(\partial _{\widehat{R}}\widehat{f})\times \partial _{\alpha }%
\widehat{R}=0$ but, in general, $\partial _{\alpha }f(R)\neq 0.$

The equations of motion for our nonholonomically modified massive gravity
theory can be written%
\begin{equation}
(\partial _{\widehat{R}}\widehat{f})\widehat{\mathbf{R}}_{\mu \nu }-\frac{1}{%
2}\widehat{f}(\widehat{R})\mathbf{g}_{\mu \nu }+\mathring{\mu}^{2}\mathbf{X}%
_{\mu \nu }=M_{Pl}^{-2}\mathbf{T}_{\mu \nu },  \label{mgrfe}
\end{equation}%
where $M_{Pl}$ is the Plank mass, $\widehat{\mathbf{R}}_{\mu \nu }$ is the
Einstein tensor for a pseudo--Riemannian metric $\mathbf{g}_{\mu \nu }$ and $%
\widehat{\mathbf{D}},$ $\mathbf{T}_{\mu \nu }$ is the standard matter
energy--momentum tensor. We note that in above formulas, for $\widehat{%
\mathbf{D}}\rightarrow \nabla ,$ we get $\widehat{\mathbf{R}}_{\mu \nu
}\rightarrow R_{\mu \nu }$ with a standard Ricci tensor $R_{\mu \nu }$ for $%
\nabla .$ Such limits give rise to original modified massive gravity models
but it is important to find some generalized solutons and only at the end to
consider additional assumptions on limits and nonholonomic constraints. It
should be emphasized that the effective energy--momentum tensor $\mathbf{X}%
_{\mu \nu }$ in (\ref{mgrfe}) is defined by the potential of graviton $%
\mathcal{U}=\mathcal{U}_{2}+\alpha _{3}\mathcal{U}_{3}+\alpha _{4}\mathcal{U}%
_{4},$ where $\alpha _{3}$ and $\alpha _{4}$ are free parameters. The values
$\mathcal{U}_{2},\mathcal{U}_{3}$ and $\mathcal{U}_{4}$ are certain
polynomials on traces of some other polynomials of a matrix $\mathcal{K}%
_{\mu }^{\nu }=\delta _{\mu }^{\nu }-\left( \sqrt{g^{-1}\Sigma }\right)
_{\mu }^{\nu }$ for a tensor determined by four St\"{u}ckelberg fields $\phi
^{\underline{\mu }}$ as
\begin{equation}
\Sigma _{\mu \nu }=\partial _{\mu }\phi ^{\underline{\mu }}\partial _{\nu
}\phi ^{\underline{\nu }}\eta _{\underline{\mu }\underline{\nu }},
\label{bm}
\end{equation}%
when $\eta _{\underline{\mu }\underline{\nu }}=(1,1,1,-1).$ A series of
arguments presented in \cite{kobayashi} (geometrically, we cand consider
corresponding nonholonomic frame transforms and nonholonomic variables)
prove that the parameter choice $\alpha _{3}=(\alpha -1)/3,\alpha
_{4}=(\alpha ^{2}-\alpha +1)/12$ is useful for avoiding potential ghost
instabilities. By frame transforms, we can fix
\begin{equation}
\mathbf{X}_{\mu \nu }=\alpha ^{-1}\mathbf{g}_{\mu \nu }.  \label{cosmconst}
\end{equation}

By explicit computations, we can prove that for configurations (\ref%
{cosmconst}), de Sitter solutions with effective cosmological constant are
possible, for instance, for ansatz of PG type,%
\begin{equation}
ds^{2}=U^{2}(r,t)[dr+\epsilon \sqrt{f(r,t)}dt]^{2}+\widetilde{\alpha }%
^{2}r^{2}(d\theta ^{2}+\sin ^{2}\theta d\varphi ^{2})-V^{2}(r,t)dt^{2}.
\label{pgm}
\end{equation}%
In above formula, there are used spherical coordinates labelled in the form $%
u^{\beta }=(x^{1}=r,x^{2}=\theta ,y^{3}=\varphi ,y^{4}=t),$ when the
function $f$\ takes non--negative values and the constant $\widetilde{\alpha
}=\alpha /(\alpha +1)$ and $\epsilon =\pm 1.$ We can consider bimetric
configurations (determined as soltuons of the system \ref{mgrfe}), \ref{bm})
and (\ref{cosmconst})) with St\"{u}ckelberg fields parameterized in the
unitary gauge as $\phi ^{\underline{4}}=t$ and $\phi ^{\underline{1}}=r%
\widehat{n}^{\underline{1}},\phi ^{\underline{2}}=r\widehat{n}^{\underline{2}%
},\phi ^{\underline{3}}=r\widehat{n}^{\underline{3}},$ where a three
dimensional (3--d) unit vector is defined as $\widehat{n}=(\widehat{n}^{%
\underline{1}}=\sin \theta \cos \varphi ,\widehat{n}^{\underline{2}}=\sin
\theta \sin \varphi ,\widehat{n}^{\underline{3}}=\cos \theta ).$

First we note that any PG metric of type (\ref{pgm}) defines solutions both
in GR and in MGT. For instance, we can extract the de Sitter solution, in
the absence of matter, and obtain standard cosmological equations with FLRW
metric, for a perfect fluid source%
\begin{equation}
T_{\mu \nu }=\left[ \rho (t)+p(t)\right] u_{\mu }u_{\nu }+p(t)g_{\mu \nu },
\label{memt}
\end{equation}%
where $u_{\mu }=(0,0,0,-V)$ can be reproduced for the effective cosmological
constant $\ ^{eff}\lambda =\mathring{\mu}^{2}/\alpha .$ Secondly, it is also
possible to express metrics of type (\ref{pgm}) in a familiar cosmological
FLRW form (see formulas (23), (24) and (27) in \cite{kobayashi}).

Finally, in this section, we note that off--diagonal deformations of such
solutions can be constructed for different classes of zero graviton mass
and/or massive theories and various f--modifications, see examples in \cite%
{vadm1} and \cite{bhnh}.

\section{Generating off--diagonal cosmological solutions}

\label{s3}

We now make a crucial assumption that our Universe can be described by
inhomogeneous cosmological metrics are with Killing symmetry on $\partial
_{3}=\partial _{\varphi }$ and, in general, can not be diagonalized by
coordinate transforms. As a matter of principle, we can consider
dependencies on all spacetime coordinates but this require more cumbersome
computations, see examples in Refs. \cite{vadm1}. Inhomogeneities and
anisotropies can be very small, but it is important to find certain classes
of general solutions for generic nonlinear systems and impose at the end
certain homogeneity and high symmetry conditions by selecting corresponding
subclasses of generation and integration functions. Up to general classes of
frame transforms, we can consider the ansatz
\begin{eqnarray}
ds^{2} &=&\eta _{1}(r,\theta )\mathring{g}_{1}(r)dr^{2}+\eta _{2}(r,\theta )%
\mathring{g}_{2}(r)d\theta ^{2}+ \omega ^{2}(r,\theta ,\varphi ,t)\{\eta
_{3}(r,\theta ,t) \times  \notag \\
&&\mathring{h}_{3}(r,\theta )[d\varphi +n_{i}(r,\theta )dx^{i}]^{2}+\eta
_{4}(r,\theta ,t)\mathring{h}_{4}(r,\theta ,t)[dt+(w_{i}(x^{k},t)+\mathring{w%
}_{i}(x^{k}))dx^{i}]^{2}\}.  \label{offdans}
\end{eqnarray}
The values $\eta _{\alpha }$ are called "polarization" functions, where $%
\omega $ is the so--called "vertical", v, conformal factor.

For metrics (\ref{offdans}), we can consider off--diagonal N--coefficients
labelled $N_{i}^{a}(x^{k},y^{4}),$ where (for parameterizations
corresponding to this class of ansatz)
\begin{equation*}
N_{i}^{3}=n_{i}(r,\theta )\mbox{ and }N_{i}^{4}=w_{i}(x^{k},t)+\mathring{w}%
_{i}(x^{k}).
\end{equation*}%
The data for the "primary" metric are
\begin{eqnarray}
\mathring{g}_{1}(r) &=&U^{2}-\mathring{h}_{4}(\mathring{w}_{1})^{2},%
\mathring{g}_{2}(r)=\widetilde{\alpha }^{2}r^{2},\mathring{h}_{3}=\widetilde{%
\alpha }^{2}r^{2}\sin ^{2}\theta ,\mathring{h}_{4}=\sqrt{|fU^{2}-V^{2}|},
\notag \\
\mathring{w}_{1} &=&\epsilon \sqrt{f}U^{2}/\mathring{h}_{4},\mathring{w}%
_{2}=0,\mathring{n}_{i}=0,  \label{pmdata}
\end{eqnarray}%
when the coordinate system is such way fixed that the values $f,U,V$ in (\ref%
{pgm}) result in a coefficient $\mathring{g}_{1}$ depending only on $r$.

Using nonholonomic frame transforms, we can parameterize the energy momentum
sources (\ref{memt}) and effective (\ref{cosmconst}) in the form
\begin{equation}
\Upsilon _{\beta }^{\alpha }=\frac{1}{M_{Pl}^{2}(\partial _{\widehat{R}}%
\widehat{f})}(\mathbf{T}_{\beta }^{\alpha }+\alpha ^{-1}\mathbf{X}_{\beta
}^{\alpha })=\frac{1}{M_{Pl}^{2}(\partial _{\widehat{R}}\widehat{f})}(\
^{m}T+\alpha ^{-1})\delta _{\beta }^{\alpha }=(\ ^{m}\Upsilon +\ ^{\alpha
}\Upsilon )\delta _{\beta }^{\alpha },  \label{effectsourc}
\end{equation}
for constant values $\ ^{m}\Upsilon :=M_{Pl}^{-2}(\partial _{\widehat{R}}%
\widehat{f})^{-1}\ ^{m}T$ and $\ ^{\alpha }\Upsilon =M_{Pl}^{-2}(\partial _{%
\widehat{R}}\widehat{f})^{-1}\alpha ^{-1},$ with respect to N--adapted
frames (\ref{nframes}). In general, such sources are not diagonal and may
depend on all spacetime coordinates. Our assumption is that we prescribe a
distribution with one Killing symmetry in a moment of time and then find
further evolution with respect to certain classes of nonholonomic frames. We
emphasize that fixing such N--adapted parameterizations we can decouple the
gravitational filed equations in MGTs and construct exact solutions in
explicit form.

Let us outline in brief the decoupling property of the gravitational and
matter field equations in GR and various generalizations/ modifications
studied in details in Refs. \cite{vadm1}. That anholonomic frame deformation
method (AFDM) can be applied for decoupling, and constructing solutions of
the MGT field equations (\ref{mgrfe}) with any effective source
parameterized in the form (\ref{effectsourc}), see details in Refs. \cite%
{vadm1}).

We label the target off--diagonal metrics as $\mathbf{g}=(g_{i}=\eta _{i}%
\mathring{g}_{i},h_{a}=\eta _{a}\mathring{h}_{a},N_{j}^{a})$ with
coefficients determined by ansatz (\ref{offdans}). In these formulas, there
is not summation on repeating indices in this formula. We shall use brief
denotations for partial derivatives: $\partial _{1}\psi =\psi ^{\bullet
},\partial _{2}\psi =\psi ^{\prime },\partial _{3}\psi =\psi ^{\diamond }$
and $\partial _{4}\psi =\psi ^{\ast }.$ Computing the N--adapted
coefficients of the Ricci tensors, when  $\widehat{R}_{1}^{1}=\widehat{R}%
_{2}^{2},\widehat{R}_{3}^{3}=\widehat{R}_{4}^{4},\widehat{R}_{3k}$ and $%
\widehat{R}_{4k}$  are not trivial, we write (\ref{mgrfe}) as a system of
nonlinear partial differential equations (PDE):%
\begin{eqnarray}
\psi ^{\bullet \bullet }+\psi ^{\prime \prime } &=&2(\ ^{m}\Upsilon +\
^{\alpha }\Upsilon ),  \label{mgfeq1} \\
\phi ^{\ast }h_{3}^{\ast } &=&2h_{3}h_{4}(\ ^{m}\Upsilon +\ ^{\alpha
}\Upsilon ),  \notag \\
n_{i}^{\ast \ast }+\gamma n_{i}^{\ast } &=&0,\beta w_{i}-\alpha _{i}=0,\
\notag \\
\partial _{k}\omega &=&n_{k}\omega ^{\diamond }+w_{k}\omega ^{\ast },
\label{conf2}
\end{eqnarray}%
for%
\begin{equation}
\phi =\ln \left\vert \frac{h_{3}^{\ast }}{\sqrt{|h_{3}h_{4}|}}\right\vert
,\gamma :=(\ln \frac{|h_{3}|^{3/2}}{|h_{4}|})^{\ast },\ \ \alpha _{i}=\frac{%
h_{3}^{\ast }}{2h_{3}}\partial _{i}\phi ,\ \beta =\frac{h_{3}^{\ast }}{2h_{3}%
}\phi ^{\ast },  \label{c1}
\end{equation}
In above formulas, we consider the system of coordinates and polarization
functions are fixed for configurations with $g_{1}=g_{2}=e^{\psi (x^{k})}$
and nonzero values $\phi ^{\ast }$ and $h_{a}^{\ast }.$

We can extract solutions for the LC--configurations with zero torsion if the
coefficients of metrics are subjected to additional conditions:
\begin{equation}
w_{i}^{\ast }=\mathbf{e}_{i}\ln \sqrt{|\ h_{4}|},\mathbf{e}_{i}\ln \sqrt{|\
h_{3}|}=0,\partial _{i}w_{j}=\partial _{j}w_{i}\mbox{ and }n_{i}^{\ast }=0.
\label{lccond}
\end{equation}

Step by step, the system of nonlinear PDE (\ref{mgfeq1})--(\ref{lccond}) can
be integrated in general forms for any $\omega $ constrained by a system of
linear first order equations (\ref{conf2}), see details in \cite{vadm1}. The
explicit solutions are given by quadratic elements%
\begin{equation}
ds^{2}=e^{\psi (x^{k})}[(dx^{1})^{2}+(dx^{2})^{2}]+\frac{\Phi ^{2}\omega ^{2}%
}{4\ (\ ^{m}\Upsilon +\ ^{\alpha }\Upsilon )}\mathring{h}_{3}[d\varphi
+\left( \partial _{k}\ n\right) dx^{k}]^{2}-\frac{(\Phi ^{\ast })^{2}\omega
^{2}}{(\ ^{m}\Upsilon +\ ^{\alpha }\Upsilon )\Phi ^{2}}\mathring{h}%
_{4}[dt+(\partial _{i}\ \widetilde{A})dx^{i}]^{2}.  \label{nvlcmgs}
\end{equation}%
for any
\begin{equation*}
\Phi =\check{\Phi},(\partial _{i}\check{\Phi})^{\ast }=\partial _{i}\check{%
\Phi}^{\ast },w_{i}+\mathring{w}_{i}=\partial _{i}\check{\Phi}/\check{\Phi}%
^{\ast }=\partial _{i}\ \widetilde{A}.
\end{equation*}
We can construct exact solutions even such conditions are not satisfied,
i.e. the zero torsion conditions are not stated or there are given certain
sources in non--explicit form. So, the AFDM can be applied to generate both
off--diagonal metrics and nonholonomically induced torsions. There are
various physical arguments for what type of generating/ integration
functions and sources we have to chose in order to construct realistic
scenarios for Universe acceleration and observable dark energy/ matter
effects.

Coming back to the properties of general solutions, we note that we can
generate new classes of solutions for arbitrary nontrivial sources, $\
^{m}\Upsilon +\ ^{\alpha }\Upsilon \neq 0,$ and generating functions, $\Phi
(x^{k},t):=e^{\phi }$ and $n_{k}=\partial _{k}n(x^{i}).$ Resulting target
metrics are generic off--diagonal and can not be diagonalized via coordinate
transforms in a finite spacetime region because, in general, the anholonomy
coefficients $W_{\alpha \beta }^{\gamma } $ for (\ref{nframes}) are not zero
(we can check by explicit computations). The polarization $\eta $--functions
from (\ref{nvlcmgs}) are
\begin{equation}
\eta _{1}=e^{\psi }/\mathring{g}_{1},\eta _{2}=e^{\psi }/\mathring{g}%
_{2},\eta _{3}=\Phi ^{2}/4\ (\ ^{m}\Upsilon +\ ^{\alpha }\Upsilon ),\eta
_{4}=(\Phi ^{\ast })^{2}/(\ ^{m}\Upsilon +\ ^{\alpha }\Upsilon )\Phi ^{2}.
\label{polarf}
\end{equation}
We conclude that prescribing any generating functions $\check{\Phi}(r,\theta
,t),$ $n(r,\theta ),$ $\omega (r,\theta ,\varphi ,t)$ and sources $\
^{m}\Upsilon ,\ ^{\alpha }\Upsilon $ and then computing $\check{A}(r,\theta
,t),$ we can transform any PG (and, similarly, FLRW) metric $\mathbf{%
\mathring{g}}=(\mathring{g}_{i},\mathring{h}_{a},\mathring{w}_{i},\mathring{n%
}_{i})$ in MGT and/or GR into new classes of generic off--diagonal exact
solutions depending on all spacetime coordinates. Such metrics define
Einstein manifolds in GR with effective cosmological constants determined by
$\ ^{m}\Upsilon +\ ^{\alpha }\Upsilon .$ With respect to N--adapted frames (%
\ref{nframes}) the coefficients of metric encode contributions from massive
gravity, determined by $\ ^{\alpha }\Upsilon ,$ and matter fields, included
in $\ ^{m}\Upsilon .$

We also note that it is possible to provide an "alternative" treatment of (%
\ref{nvlcmgs}) as exact solutions in MGT. \ In such a case, we have to
define and analyze the properties of fiducial St\"{u}ckelberg fields $\phi ^{%
\underline{\mu }}$ and the corresponding bimetric structure resulting in
target solutions $\mathbf{g}=(g_{i},h_{a},N_{j}^{a}):$

Let us analyze the primary configurations related to
\begin{equation*}
\mathring{\phi}^{\underline{\mu }}=(\mathring{\phi}^{\underline{i}}=a(\tau
)\rho \widetilde{\alpha }^{-1}\widehat{n}^{\underline{i}},\mathring{\phi}^{%
\underline{3}}=a(\tau )\rho \widetilde{\alpha }^{-1}\widehat{n}^{\underline{3%
}},\mathring{\phi}^{\underline{4}}=\tau \kappa ^{-1}),
\end{equation*}
when the corresponding prime PG--metric $\mathbf{\mathring{g}}$ is taken in
FLRW form
\begin{equation*}
ds^{2}=a^{2}(d\rho ^{2}/(1-K\rho ^{2})+\rho ^{2}(d\theta ^{2}+\sin
^{2}\theta d\varphi ^{2}))-d\tau ^{2}.
\end{equation*}
A corresponding fiducial tensor (\ref{bm}) is computed
\begin{equation*}
\mathring{\Sigma}_{\underline{\mu }\underline{\nu }}du^{\underline{\mu }}du^{%
\underline{\nu }}=\frac{a^{2}}{\tilde{\alpha}^{2}}[d\rho ^{2}+\rho
^{2}(d\theta ^{2}+\sin ^{2}\theta d\varphi ^{2})+2H\rho d\rho d\tau -(\frac{%
\tilde{\alpha}^{2}}{\kappa ^{2}a^{2}}-H^{2}\rho ^{2})d\tau ^{2}],
\end{equation*}%
where the coefficients and coordinates are re--defined in the form $%
r\rightarrow \rho =\widetilde{\alpha }r/a(\tau )$ and $t\rightarrow \tau
=\kappa t,$ for $K=0,\pm 1;\kappa $ is an integration constant; $H:=d\ln
a/d\tau $ and the local coordinates are parameterized in the form $x^{%
\underline{1}}=\rho ,x^{\underline{2}}=\theta ,y^{\underline{3}}=\varphi ,y^{%
\underline{4}}=\tau .$

At the next step, for a target metric $\mathbf{g}={\mathbf{g}_{\alpha \beta }%
}$ and frames $\mathbf{e}_{\alpha }^{\ }=\mathbf{e}_{\alpha }^{\ \underline{%
\alpha }}\partial _{\underline{\alpha }},$ we write
\begin{equation*}
\mathbf{g}_{\alpha \beta }=\mathbf{e}_{\alpha }^{\ \underline{\alpha }}%
\mathbf{e}_{\beta }^{\ \underline{\beta }}\eta _{\underline{\alpha }%
\underline{\beta }}=\left[
\begin{array}{cc}
g_{ij}+N_{i}^{a}N_{j}^{b}g_{ab} & N_{i}^{a}h_{ab} \\
N_{i}^{a}h_{ab} & h_{ab}%
\end{array}%
\right] ,\mbox{ for }\mathbf{e}_{\alpha }^{\ \underline{\alpha }}=\left[
\begin{array}{cc}
\mathbf{e}_{i}^{\ \underline{i}} & N_{i}^{b}\mathbf{e}_{b}^{\ \underline{a}}
\\
0 & \mathbf{e}_{a}^{\ \underline{a}}%
\end{array}%
\right] .
\end{equation*}%
The values
\begin{equation*}
g_{ij}=\mathbf{e}_{i}^{\ \underline{\alpha }}\mathbf{e}_{j}^{\ \underline{%
\beta }}\eta _{\underline{\alpha }\underline{\beta }}=e^{\psi }\delta
_{ij}=diag[\eta _{i}\mathring{g}_{i}],h_{ab}=\mathbf{e}_{i}^{\ \underline{%
\alpha }}\mathbf{e}_{j}^{\ \underline{\beta }}\eta _{\underline{\alpha }%
\underline{\beta }}=diag[\eta _{a}\mathring{h}_{a}],N_{i}^{3}=\partial _{k}\
n,N_{i}^{4}=\partial _{k}\ \widetilde{A}
\end{equation*}%
are related algebraically to data (\ref{polarf}) resulting in off--diagonal
solutions (\ref{nvlcmgs}). Then, to work out the the "target" St\"{u}%
ckelberg fields we compute $\phi ^{\mu ^{\prime }}=\mathbf{e}_{\ \underline{%
\mu }}^{\mu ^{\prime }\ }\phi ^{\underline{\mu }}$ with $\mathbf{e}_{\
\underline{\mu }}^{\mu ^{\prime }\ }$ being inverse to $\mathbf{e}_{\alpha
}^{\ \underline{\alpha }},$ and the fiducial tensor
\begin{equation*}
\Sigma _{\alpha \beta }=(\mathbf{e}_{\alpha }^{\ }\phi ^{\underline{\mu }})(%
\mathbf{e}_{\beta }^{\ }\phi ^{\underline{\nu }})\eta _{\underline{\mu }%
\underline{\nu }}=\mathbf{e}_{\alpha }^{\ \underline{\alpha }}\mathbf{e}%
_{\beta }^{\ \underline{\beta }}\Sigma _{\underline{\alpha }\underline{\beta
}}.
\end{equation*}
If the prime value $\mathring{\Sigma}_{\underline{\mu }\underline{\nu }}$
carries information about two constants $\kappa $ and $\widetilde{\alpha },$
a target tensor $\Sigma _{\mu \nu }$ is associated to off--diagonal
solutions and encodes data about generating and integration functions and
via superpositions on possible Killing symmetries, on various integration
constants. Similar constructions were elaborated for holonomic and
nonholonomic configurations in GR, see \cite{geroch}.

In the framework of MGT, two cosmological solutions $\mathbf{\mathring{g}}$
and $\mathbf{g}$ related by nonholonomic deformations\footnote{%
involving not only frame transforms but also deformation of the linear
connection structure when at the end there are imposed additional
constraints for zero torsion} are characterised respectively by two
invariants
\begin{equation*}
\mathring{I}^{\underline{\alpha }\underline{\beta }}=\mathring{g}^{\alpha
\beta }\partial _{\alpha }\mathring{\phi}^{\underline{\alpha }}\partial
_{\beta }\mathring{\phi}^{\underline{\beta }}\mbox{ and }\mathbf{I}^{\alpha
\beta }=\mathbf{g}^{\alpha \beta }\mathbf{e}_{\alpha }\phi ^{\underline{%
\alpha }}\mathbf{e}_{\beta }\phi ^{\underline{\beta }}.
\end{equation*}%
The tensor $\mathring{I}^{\underline{\alpha }\underline{\beta }}$ does not
contain singularities because there are not coordinate singularities on
horizon for PG metrics. It should be emphasized that the symmetry of $\Sigma
_{\mu \nu }$ is not the same as that of $\mathring{\Sigma}_{\underline{\mu }%
\underline{\nu }}$ and the singular behaviour of $\mathbf{I}^{\alpha \beta }$
depends on the class of generating and integration functions we chose of
constructing a target solution $\mathbf{g}$.

In GR, MGTs and/or Einstein--Finsler gravity theories \cite{vadm1},
off--cosmological solutions of type (\ref{nvlcmgs}) were found to
generalized various models of Biachi, Kasner, G\"{o}del and other universes.
There are known locally anisotropic black hole and wormhole, in general,
with solitonic background solutions, see \cite{bhnh,vacarsolitonhier}. For
instance, Bianchi type anisotropic cosmological metrics are generated if we
impose corresponding Lie algebra symmetries on metrics. It was emphasized in
\cite{kobayashi} that "any PG--type solution in general relativity (with a
cosmological constant) is also a solution to massive gravity." Such a
conclusion can be extended to a large class of generic off--diagonal
cosmological solutions generated by effective cosmological constants but it
is not true, for instance, if we consider nonholonomic deformations with
nonholonomically induced torsion like in metric compatible Finsler theories.

Finally, we note that the analysis of cosmological perturbations around an
off--diagonal cosmological background is not trivial because the fiducial
and reference metrics do not respect the same symmetries. Nevertheless,
fluctuations around de Sitter backgrounds seem to have a decoupling limit
which implies that one can avoid potential ghost instabilities if the
parameter choice is considered both for diagonal and off--diagonal
cosmological solutions, see details in \cite{fluct}. This special choice
also allows us to have a structure $X_{\mu \nu }\sim g_{\mu \nu }$ at list
in N--adapted frames when the massive gravity effects can be approximated by
effective cosmological constants and exact solutions in MGT which are also
solutions in GR.

\section{Examples of off--diagonal solutions with solitonic configurations}

\label{s4} We now consider three examples of off--diagonal cosmological
solutions with solitonic modifications in MGT and (with alternative
interpretation) GR. Two and three dimensional solitonic waves are typical
nonlinear wave configurations which can be used for generating spacetime
metrics with Killing, or non--Killing, symmetries and can be characterised
by additional parametric dependencies and solitonic symmetries. Moving
solitonic configurations can mimic various types of modified gravity dark
energy and dark matter effects with nontrivial gravitational vacuum,
polarization of constants and additional nonlinear diagonal and
off--diagonal interactions of the gravitational and matter fields.

\subsection{One soliton solutions}

We shall for simplicity work with a nonlinear radial (solitonic, with left $s
$-label) generating function
\begin{equation}
\Phi =\ ^{s}\check{\Phi}(r,t)=4\arctan e^{q\sigma (r-vt)+q_{0}}
\label{1solgf}
\end{equation}%
and $\omega =1,$ we construct a metric
\begin{equation}
\mathbf{ds}^{2} =e^{\psi (r,\theta )}(dr^{2}+\ d\theta ^{2})+~\frac{\ ^{s}%
\check{\Phi}^{2}}{4\ (\ ^{m}\Upsilon +\ ^{\alpha }\Upsilon )}\mathring{h}%
_{3}(r,\theta )d\varphi ^{2} -\frac{(\partial _{t}\ ^{s}\check{\Phi})^{2}}{%
(\ ^{m}\Upsilon +\ ^{\alpha }\Upsilon )\ ^{s}\check{\Phi}^{2}}\mathring{h}%
_{4}(r,t)[dt+(\partial _{r}\ \widetilde{A})dr]^{2},  \label{offdsol1}
\end{equation}%
In this metric, for simplicity, we fixed $n(r,\theta )=0$\ and consider that
$\widetilde{A}(r,t)$ is defined as a solution of $\ ^{s}\check{\Phi}%
^{\bullet }/\ ^{s}\check{\Phi}^{\ast }=\partial _{r}\ \widetilde{A}$ and $%
\mathring{h}_{a}$ are given by PG--data (\ref{pmdata}). The generating
function (\ref{1solgf}), where $\sigma ^{2}=(1-v^{2})^{-1}$ for constants $%
q,q_{0},v,$ is just a 1--soliton solution of the sine--Gordon equation
\begin{equation*}
\ ^{s}\check{\Phi}^{\ast \ast }-\ ^{s}\check{\Phi}^{\bullet \bullet }+\sin \
^{s}\check{\Phi}=0.
\end{equation*}
For any class of small polarizations with $\eta _{a}\sim 1),$ we can
consider that the source $(\ ^{m}\Upsilon +\ ^{\alpha }\Upsilon )$ is
polarized by $\ ^{s}\check{\Phi}^{-2}$ when $h_{3}\sim \mathring{h}_{3}$ and
$h_{4}\sim \mathring{h}_{4}(\ ^{s}\check{\Phi}^{\ast })^{2}/\ ^{s}\check{\Phi%
}^{-4}$ with an off--diagonal term $\partial _{r}\ \widetilde{A}$ resulting
in a stationary solitonic universe. If we consider that $(\partial _{%
\widehat{R}}\widehat{f})^{-1}=\ ^{s}\check{\Phi}^{-2}$ in (\ref{effectsourc}%
), we can model $\widehat{f}$--interactions of type (\ref{act}) via
off--diagonal interactions and "gravitational polarizations".

In absence of matter, $\ ^{m}\Upsilon =0,$ the off--diagonal cosmology is
completely determined by $\ ^{\alpha }\Upsilon $ when $\ ^{s}\check{\Phi}$
transforms $\mathring{\mu}$ into an anisotropically polarized/variable mass
of solitonic waves. Such configurations can be modelled if $\ ^{m}\Upsilon
\ll \ ^{\alpha }\Upsilon .$ If $\ ^{m}\Upsilon \gg \ ^{\alpha }\Upsilon ,$
we generate cosmological models determined by distribution off matter fields
when contributions from massive gravity are with small anisotropic
polarization. For a class of nonholonomic constraints on $\Phi $ and $\psi $
(which may be not of solitonic type), when solutions (\ref{offdsol1}) are of
type (\ref{offdans}) with $\eta _{\alpha }\sim 1$ and $n_{i},w_{i}\sim 0,$
we approximate PG--metrics of type (\ref{pgm}).

Hence, by an appropriate choice of generating functions and sources, we can
model equivalently modified gravity effects, massive gravity contributions
or matter field configurations in GR and MGT interactions. For well defined
conditions, such configurations can be studied in the framework of some
classes of off--diagonal solutions in Einstein gravity with effective
cosmological constants.

\subsection{Three dimensional solitonic anistoropic waves}

More sophisticate nonlinear gravitational and matter filed interactions can
be modeled both in MGTs and GR if we consider more general classes of
solutions of effective Einstein equations.

For instance, off--diagonal solitonic metrics with one Killing symmetry can
be generated, for instance, if we take instead of (\ref{1solgf}) a
generating functions $\ ^{s}\check{\Phi}(r,\theta ,t)$ which is a solution
of the Kadomtsev--Petivashvili, KdP, equations \cite{kadom},%
\begin{equation*}
\pm \ ^{s}\check{\Phi}^{\prime \prime }+(\ ^{s}\check{\Phi}^{\ast }+\ ^{s}%
\check{\Phi}\ \ ^{s}\check{\Phi}^{\bullet }+\epsilon \ ^{s}\check{\Phi}%
^{\bullet \bullet \bullet })^{\bullet }=0,
\end{equation*}%
when solutions induce certain anisotropy on $\theta .$\footnote{%
In a similar form, we can construct various types of vacuum gravitational
2-d and 3-d configurations characterized by solitonic hierarchies and
related bi--Hamilton structures, for instance, of KdP equations with
possible mixtures with solutions for 2-d and 3-d sine--Gordon equations etc,
see details in Ref. \cite{vacarsolitonhier}.} In the dispersionless limit $%
\epsilon \rightarrow 0,$ we can consider that the solutions are independent
on $\theta $ and determined by Burgers' equation
\begin{equation*}
\ ^{s}\check{\Phi}^{\ast }+\ ^{s}\check{\Phi}\ \ ^{s}\check{\Phi}^{\bullet
}=0.
\end{equation*}

Such solutions can be parameterized and treated similarly to (\ref{offdsol1}%
) but with, in general, a nontrivial term $(\partial _{\theta }\ \widetilde{A%
})d\theta $ after $\mathring{h}_{4},$ when $\ ^{s}\check{\Phi}^{\bullet }/\
^{s}\check{\Phi}^{\ast }=\ \widetilde{A}^{\bullet }$ and $\ ^{s}\check{\Phi}%
^{\prime }/\ ^{s}\check{\Phi}^{\ast }=\ \widetilde{A}^{\prime }.$ Similar
metrics were constructed for Dirac spinor waves and solitons in anisotropic
Taub-NUT spaces and in five dimensional brane gravity which can be encoded
and classified by corresponding solitonic hierarchies and geometric
invariants, see \cite{vacarsolitonhier}. Here we proved that AFDM can
extended to generate inhomogeneous off--diagonal cosmological solitonic
solutions in varios MGTs.

\subsection{ Solitonic waves for a nontrivial vertical conformal $v$--factor}

The cosmological solutions we look are also with three dimensional solitons
but for a $v$--factor as in (\ref{offdans}). For instance, we consider
solitons of KdP type, when $\omega =\check{\omega}(r,\varphi ,t),$ when $%
x^{1}=r, x^{2}=\theta ,y^{3}=\varphi ,y^{4}=t,$ for
\begin{equation}
\pm \check{\omega}^{\diamond \diamond }+(\partial _{t}\check{\omega}+\check{%
\omega}\ \check{\omega}^{\bullet }+\epsilon \check{\omega}^{\bullet \bullet
\bullet })^{\bullet }=0,  \label{kdp1}
\end{equation}%
In the dispersionless limit $\epsilon \rightarrow 0,$ we can consider that
the solutions are independent on angle $\varphi $ and determined by Burgers'
equation
\begin{equation*}
\check{\omega}^{\ast }+\check{\omega}\ \check{\omega}^{\bullet }=0.
\end{equation*}
The conditions (\ref{conf2}) impose an additional constraint
\begin{equation*}
\mathbf{e}_{1}\check{\omega}=\check{\omega}^{\bullet }+w_{1}(r,\theta
,\varphi )\check{\omega}^{\ast }+n_{1}(r,\theta )\check{\omega}^{\diamond
}=0.
\end{equation*}
In the system of coordinates when $\check{\omega}^{\prime }=0,$ we can fix $%
w_{2}=0$ and $n_{2}=0.$ For any arbitrary generating function with
LC--configuration, $\check{\Phi}(r,\theta ,t),$ we construct exact solutions
\begin{equation}
\mathbf{ds}^{2} =e^{\psi (r,\theta )}(dr^{2}+\ d\theta ^{2})+~\frac{\check{%
\Phi}^{2}\check{\omega}^{2}}{4\ (\ ^{m}\Upsilon +\ ^{\alpha }\Upsilon )}%
\mathring{h}_{3}(r,\theta )d\varphi ^{2} -\frac{(\partial _{t}\ \check{\Phi}%
)^{2}\check{\omega}^{2}}{(\ ^{m}\Upsilon +\ ^{\alpha }\Upsilon )\ \check{\Phi%
}^{2}}\mathring{h}_{4}(r,t)[dt+(\partial _{r}\ \widetilde{A})dr]^{2},
\label{offdsol3}
\end{equation}%
which are generic off--diagonal and depend on all spacetime coordinates.
Such stationary cosmological solutions are with polarizations on two angles $%
\theta $ and $\varphi .$ Nevertheless, the character of anisotropies is
different for metrics of type (\ref{offdsol1}) and (\ref{offdsol3}). In the
third class of metrics, we obtain a Killing symmetry on $\partial _{\varphi
} $ only in the limit  $\check{\omega}\rightarrow 1,$  but in the first two
ones, such a symmetry exists generically. For (\ref{offdsol3}), the value $%
\check{\Phi}$ is not obligatory a solitonic one which can be used for
additional off--diagonal modifications of solutions and various types of
polarizations.

We can provide a physical interpretation of \ref{offdsol3} which is similar
to \ref{offdsol1} if the generating and integration functions are chosen to
satisfy the conditions $\eta _{\alpha }\sim 1$ and $n_{i},w_{i}\sim 0,$ we
approximate PG--metrics of type (\ref{pgm}).

In a particular case, we can use a conformal $v$--factor which is a
1--solitonic one, i.e.
\begin{equation*}
\check{\omega}\rightarrow \omega (r,t)=4\arctan e^{q\sigma (r-vt)+q_{0}},
\end{equation*}
where $\sigma ^{2}=(1-v^{2})^{-1}$ and constants $q,q_{0},v,$ defines a
1--soliton solution of the sine--Gordon equation
\begin{equation*}
\omega ^{\ast \ast }-\omega ^{\bullet \bullet }+\sin \omega =0.
\end{equation*}
Such a soliton propagates in time along the radial coordinate.

We conclude that solitonic waves may mimic both particle type configurations
as dark matter and encode certain hidden dark energy and off--diagonal
gavitational and matter field interactions.

\section{Reconstruction mechanism for off--cosmological solutions}

\label{s5} We now consider a reconstruction mechanism with distinguished
off--diagonal cosmological effects \cite{vadm1} by generalizing some methods
elaborated for $f(R)$ gravity in \cite{odints}. The main idea was to present
the MGT actions (with zero or non-zero gravitational mass) as sums of
actions in GR and certain effective ideal fluid contributions with
parameters defined by nontrivial $\mathring{\mu}$ and $f$--deformations. The
reconstruction method was developed for such formulations which lead to
cosmology with cyclic evolution. Then, it was proven that the ekpyrotic
scenario may be also realized for MGTs and that it is possible to
reconstruct models of $f(R)$ gravity which induces little rip cosmology.

In our approach, we work with cosmological generic off--diagonal metrics and
generalized connections. We can always chose such generating functions and
parameters of solutions that off--diagonal contributions are small, and the
torsion is constrained to be zero, beginning a certain fixed moment of time.
Nevertheless, we can not neglect for such classes of solutions possible
nonlinear effects determined by  generating functions and effective sources 
resulting in diagonalized modifications.  MGTs were elaborated as
realistic alternatives for unified description of inflation with dark energy
when cosmological scenarios encode information on various massive and
zero--mass gravitational modes.  A crucial question is if and how such
constructions have to be elaborated for generic off--diagonal cosmological
spaces. This is not only a  geometric problem for  generalizing the
reconstruction formalism for inhomogeneous and locally anisotropic
cosmological theories. It is connected to a very important question on
equivalent modeling of cosmological scenarios for different MGTs in the
framework of GR with nonholonomic and off--diagonal nonlinear interactions.

Any cosmological solution in massive, MGT and/or GR parameterized in a form (%
\ref{offdans}) (in particular, as (\ref{offdsol1}) and (\ref{offdsol3})) can
be encoded into an effective functional (\ref{mgrfunct}) when
\begin{equation*}
\widehat{f}-\frac{\mathring{\mu}^{2}}{4}\mathcal{U}=f(\widehat{R}),\widehat{R%
}_{\mid \widehat{\mathbf{D}}\rightarrow \nabla }=R.
\end{equation*}%
This allows us to work as in MGT; the conditions $\partial _{\alpha }f(%
\widehat{R})=0$ if $\widehat{R}=const$ simplify substantially the
computations. Nevertheless, the contributions of the parameter $\mathring{\mu}$
and effective potential $\mathcal{U}$ are included as functional
dependencies of $f$ and $\widehat{R}.$For $\widehat{\mathbf{D}}\rightarrow
\nabla ,$ we get nonlinear modifications both in diagonal and off--diagonal
terms of cosmological metrics.

The starting point of our approach is to consider a prime flat FLRW like
metric
\begin{equation*}
ds^{2}=a^{2}(t)[(dx^{1})^{2}+(dx^{2})^{2}+(dy^{3})^{2}]-dt^{2},
\end{equation*}%
where $t$ is the cosmological time. In order to extract a monotonically
expanding and periodic cosmological scenario, we parameterize $\ln
|a(t)|=H_{0}t+\tilde{a}(t)$ for a periodic function $\tilde{a}(t+\tau )=\
^{1}a\cos (2\pi t/\tau ),$ where $0<\ ^{1}a<H_{0}.$ Our goal is to prove
that such a behavior is encoded into off--diagonal solutions of type (\ref%
{offdsol1})--(\ref{offdsol3}).

We write FLRW like equations with respect to N--adapted (moving) frames (\ref%
{nframes}) for a generalized Hubble function $H,$%
\begin{equation*}
3H^{2}=8\pi \rho \mbox{ and }3H^{2}+2\mathbf{e}_{4}H=-8\pi p.
\end{equation*}%
Using variables with $\partial _{\alpha }f(\widehat{R})_{|\widehat{R}%
=const}=0,$ we can consider a function $H(t)$ when $\mathbf{e}_{4}H=\partial
_{t}H=H^{\ast }.$ It should be noted that the approximation $\mathbf{e}%
_{4}\rightarrow \partial _{t}$ is considered "at the end" (after a class of
off-diagonal solutions was found for a necessary type connection, which
allowed to decouple the equations) and the generating functions and
effective sources are constrained to depend only on $t.$ The energy--density
and pressure of an effective perfect fluid are computed
\begin{eqnarray}
\rho  &=&(8\pi )^{-1}[(\partial _{R}f)^{-1}(\frac{1}{2}f(R)+3H\mathbf{e}%
_{4}(\partial _{R}f))-3\mathbf{e}_{4}H]  \label{rhomg} \\
&=&(8\pi )^{-1}[\partial _{\widehat{R}}\ln \sqrt{|\widehat{f}|}-3H^{\ast
}]=(8\pi )^{-1}[\partial _{\widehat{R}}\ln \sqrt{|\frac{\mathring{\mu}^{2}}{4%
}\mathcal{U}+f(\widehat{R})|}-3H^{\ast }],  \notag \\
&&  \notag \\
p &=&-(8\pi )^{-1}[(\partial _{R}f)^{-1}(\frac{1}{2}f(R)+2H\mathbf{e}%
_{4}(\partial _{R}f)+\mathbf{e}_{4}\mathbf{e}_{4}(\partial _{R}f))+\mathbf{e}%
_{4}H]  \notag \\
&=&(8\pi )^{-1}[\partial _{\widehat{R}}\ln \sqrt{|\widehat{f}|}+H^{\ast
}]=-(8\pi )^{-1}[\partial _{\widehat{R}}\ln \sqrt{|\frac{\mathring{\mu}^{2}}{%
4}\mathcal{U}+f(\widehat{R})|}+H^{\ast }].  \notag
\end{eqnarray}%
We emphasize that, in general, such values are defined with respect to
nonholonomic frames. The effect of  nonlinear deformations
encoding physical data $(\mathring{\mu},\mathcal{U},f,\widehat{R})$ is
preserved also in diagonal contributions for $\mathbf{e}_{4}\rightarrow
\partial _{t}.$

The equation of state, EoS, parameter for the effective dark fluid encoding
MGTs parameters is defined by
\begin{equation}
w=\frac{p}{\rho }=\frac{\widehat{f}+2H^{\ast }\partial _{\widehat{R}}%
\widehat{f}}{\widehat{f}-6H^{\ast }\partial _{\widehat{R}}\widehat{f}}=\frac{%
\frac{\mathring{\mu}^{2}}{4}\mathcal{U}+f(\widehat{R})+2H^{\ast }\partial _{%
\widehat{R}}\widehat{f}}{\frac{\mathring{\mu}^{2}}{4}\mathcal{U}+f(\widehat{R%
})-6H^{\ast }\partial _{\widehat{R}}\widehat{f}},  \label{wcosm}
\end{equation}%
when the corresponding EoS is
\begin{equation*}
p=-\rho -(2\pi )^{-1}H^{\ast }
\end{equation*}%
and $\mathcal{U}(t)$ is computed, for simplicity, for a configuration of
"target" St\"{u}ckelberg fields $\phi ^{\mu ^{\prime }}=\mathbf{e}_{\
\underline{\mu }}^{\mu ^{\prime }\ }\phi ^{\underline{\mu }}$ when a found
solution is finally modelled by generating functions with dependencies on $t.
$ The components of $\phi ^{\mu ^{\prime }}$ are computed with respect to
N--adapted frames.

Taking a generating Hubble parameter $H(t)=H_{0}t+H_{1}\sin \omega t,$ for $%
\omega =2\pi /\tau ,$ we can recover the modified action for oscillations of
off--diagonal (massive) universe (see similar details in \cite{odints}),%
\begin{equation}
f(R(t))=6\omega H_{1}\int dt[\omega \sin \omega t-4\cos \omega
t(H_{0}+H_{1}\sin \omega t)]\exp [H_{0}t+\frac{H_{1}}{\omega }\sin \omega t].
\label{recovf}
\end{equation}%
Here we note that we can not invert analytically to find in explicit form $R,
$ or any nonholonomic deformation to $\widehat{R}$ with respect to general
N--adapted frames. Nevertheless, we can prescribe any values of constants $%
H_{0}$ and $H_{1}$ and of $\omega $ and compute effective dark energy and
dark matter oscillating cosmology effects for any off--diagonal solution in
massive gravity and/or effective MGT, GR. To extract contributions of $%
\mathring{\mu}$ we can fix, for instance, $\widehat{f}(\widehat{R})=$ $%
\widehat{R}=R$ and using (\ref{act}) and (\ref{act1}) we can relate $f(R(t))$
and respective constants to certain observable data in cosmology.

The MGT theories studied in this work encode, for respective nonholonomic
constraints, the ekpyrotic scenario which can be modelled similarly to $f(R)$
gravity. A scalar field is introduced into usual ekpyrotic models in order
to reproduce a cyclic universe and such a property exists if we consider
off--diagonal solutions with massive gravity terms and/or $f$%
--modifications. The main idea is to develop the reconstruction techniques
for the scalar--tensor theory using the AFDM, with nonholonomic
off--diagonal metric and linear connection deformations. Working with
general classes of off--diagonal cosmological solutions, the problem is to
state the conditions for generating and integration functions when a
corresponding ekpyrotic scenario will "survive" for certain constraints and
in respective limits.

Let us consider a prime configuration with energy--density for pressureless
matter $\mathring{\rho}_{m},$ for radiation and anisotropies we take
respectively $\mathring{\rho}_{r}$ and $\mathring{\rho}_{\sigma }$ for
radiation and anisotropies, $\kappa $ is the spatial curvature of the
universe and a target effective energy--density $\rho $ (\ref{rhomg}). A
FLRW model can be described by
\begin{equation*}
3H^{2}=8\pi \lbrack \frac{\mathring{\rho}_{m}}{a^{3}}+\frac{\mathring{\rho}%
_{r}}{a^{4}}+\frac{\mathring{\rho}_{\sigma }}{a^{6}}-\frac{\kappa }{a^{2}}%
+\rho ].
\end{equation*}%
We generate an off--diagonal/massive gravity cosmological \ cyclic scenario
containing a contracting phase by solving the initial problems if $w>1,$ see
(\ref{wcosm}). A homogeneous and isotropic spatially flat universe is
obtained when the scale factor tends to zero and the effective $f$--terms
(massive gravity and off--diagonal contributions) dominate over the rests.
In such cases, the results are similar to those in the inflationary
scenario. For recovering (\ref{recovf}), the ekpyrotic scenario takes place
and mimic the observable universe for $t\sim \pi /2\omega $ in the effective
EoS parameter
\begin{equation*}
w\approx -1+\sin \omega t/3\omega H_{1}\cos ^{2}\omega t\gg 1.
\end{equation*}
This allows us to conclude that in massive gravity and/or using
off--diagonal interactions in GR cyclic universes can be reconstructed in
such forms that the initial, flatness and/or horizon problems can be solved.
We can compute possible locally anisotropic and inhomogeneous small
contributions for self--consistent models with nonholonomic frames.

In the diversity of off--diagonal cosmological solutions which can
constructed using above presented methods, there are cyclic ones with
singularities of the type of big bang/ crunch behaviour. This is still a
largely unexplored area both for the geometric methods of constructing exact
solutions of PDEs and recovering procedures for certain "preferred"
fundamental physical objects compatible with experimental data.  Choosing
necessary types generating and integration functions, we can avoid
singularities and elaborate models with smooth transition. Using the
possibility to generate nonholonomically constrained $f$--models with
equivalence to certain classes of solutions in massive gravity and/or
off--diagonal configurations in GR, we can study in this context, following
methods in \cite{odints,vadm1}, big and/or little rip cosmology models, when
the phantom energy--density is modelled by off-diagonal interactions. We
note that such nonsingular models for dark energy were proposed as
alternatives to $\Lambda $CDM cosmologies, \cite{littler}.  Using
corresponding classes of generating functions, we can reproduce the
sceanaros with a phantom scalar modeling a little rip in the framework of
AFDM and nonholonomic MGTs.   We omit such considerations in this article
(see a summary of such construction and further developments in \cite{vlett}).

\section{Conclusions}

\label{sconcl} In this paper we deal with new classes cosmological
off--diagonal solutions in massive gravity with flat, open and closed
spatial geometries. These solutions can be systematically constructed for
various types of modified gravity theories, MGTs, and in general relativity,
GR. We applied an advanced geometric techniques for decoupling the field
equations and constructing exact solutions in massive and zero mass $f(R)$
gravity, theories with nontrivial torsion and noholonomic constraints to GR
and possible extensions on (co) tangent Lorentz bundles. The so--called
anholonomic frame deformation method, AFDM, was elaborated during last 15
years in our works \cite{vadm1,vacarsolitonhier}, where a number of examples
of off--diagonal solutions and new applications in gravity and modern
cosmology were considered.

A very important property of such generalized classes of cosmological
solutions is that they depend, in general, on all spacetime coordinates via
generating and integration functions and constants. They describe certain
models of inhomogeneous and locally anisotropic cosmology with less clear
physical meaning and possible physical implications \cite{kourets}. After
some classes of solutions were constructed in a most general form, we can
impose at the end additional nonholonomic constraints, cosmological
approximations, extract configurations with a prescribed spacetime symmetry
and/or dependence on certain mass parameters, consider asymptotic conditions
etc. Thus, our solutions can be used for elaborating homogeneous and
isotropic cosmological models with arbitrary spatial curvature, to study
generalized Killing and non-Killing, with possible nonholonomically deformed
(super) symmetries \cite{bhnh,geroch} and to study "non--spherical" collapse
models of the formation of cosmic structure such as stars and galaxies (see
also \cite{kobayashi}).

Another aspect of the AFDM is that if we work only with cosmological
scenarios for diagonalizable metrics, there are possibilities to
discriminate the massive gravity theory form the $f$--gravity and/or GR. For
diagonal metrics depending on a time like coordinate, we can formulate
mathematical cosmology problems for certain nonlinear systems of ordinary
differential equations with general solutions depending on integration
constants. Identifying such a constant, for instance, with a graviton mass
parameter, we do not have much possibilities to mimic a number of similar
effects in GR for different MGTs. Following only such a "diagonal" approach,
we positively have to modify the GR theory in order to explain observational
data in modern cosmology and elaborate realistic quantum models of massive
gravity.

We now discuss a new and important feature of the off--diagonal anisotropic
configurations which allows us to model cosmic accelerations and massive
gravity and/or dark energy and dark matter effects as certain effective
Einstein spaces. Having integrated such system of nonlinear partial
differential equations, PDEs, for a large class of such solutions, we can
put and motivate such questions:\ May be we do not need to modify radically
the GR theory but only to extend the constructions to off--diagonal
solutions and nonholonomic systems and try to apply this in modern
cosmology? Could we explain observational data in modern cosmology via
nonlinear diagonal and/or off--diagonal interactions with non--minimal
coupling for matter and/or different phases of massive and zero mass
gravity. This is a quite complicated theoretical and experimental problem
and the main goal of this and our recent papers cited in Refs. \cite%
{vadm1,vlett} was to analyze such constructions from the viewpoint of
massive gravity theory when off--diagonal effects can be alternatively
explained to other types of gravity theories.

The reconstruction procedure for cosmological models with non--minimally
coupled scalar fields evolving on a flat FLRW background and in different
MGTs was studied in \cite{kamensh,odints}. In this work, we elaborated a
reconstruction method for the massive gravity theory which admits an
effective off--diagonal interpretation in GR and $f$--modified gravity with
cyclic and ekpyrotic universe solution. We concluded that the expansion can
be around the GR action even if we admit a nontrivial effective torsion. For
zero torsion constraints, it is possible to construct off--diagonal
cosmological models keeping the approach in the framework of the GR theory.
We further investigated how our results indicate that theories with massive
gravitons, with possible $f$--modified terms and off--diagonal interactions
may lead to more complicated scenarios of cyclic universes. Following such
nonlinear (off--diagonal) approaches, the ekpyrotic (little rip) scenario
can be realized with no need to introduce additional fields (or modifying
gravity) but only in terms of massive gravity or GR. Another interesting
constructions can be related to reconstruction scenarios of $f(R)$ and
massive gravity theories leading to little rip universes both in locally
anisotropic and isotropic variants. Finally, we note that the dark energy
for little rip models present an example of non--singular phantom cosmology.

\vskip5pt

\textbf{Acknowledgments:\ } The work is partially supported by the Program
IDEI, PN-II-ID-PCE-2011-3-0256 and performed for a corresponding associate
visiting program at CERN. The author thanks S. Capozziello, E. Guendelman, E. Elizalde, N. Mavromatos, 
M. Sami, D. Singleton, P. Stavrinos and S. Rajpoot for important
discussions, critical remarks, collaboration and substantial support.


\end{document}